\documentclass[]{aa}
\usepackage{graphicx}
\usepackage{txfonts}
\begin{document}
\title {On the potassium-rotation connection in late-type Alpha
Persei stars\thanks{Based on observations made with the Isaac Newton
Telescope, operated on the island of La Palma by the Isaac Newton
Group in the Spanish Observatorio del Roque de los Muchachos of the
Instituto de Astro{f\'\i}sica de Canarias.} }

\author {E.L. Mart{\'\i}n\inst{1,2}, A. Magazz\`u\inst{3}, R.J.
Garc{\'\i}a L\'opez\inst{1}, 
S. Randich\inst{4}, \and D. Barrado y Navascu\'es\inst{5}}

\offprints{E.L. Mart\'\i n}

\institute {
Instituto de Astrof\'\i sica de Canarias, E-38200 La Laguna, Spain\\
              \email{ege@iac.es}
\and University of Central Florida, Department of Physics, PO Box 162385, Orlando, FL 32816-2385, 
USA \\
\email{ege@physics.ucf.edu}
\and Centro Galileo Galilei, Apartado 565, E-38700 Santa Cruz de La Palma, Spain \\ 
\email{magazzu@tng.iac.es}  
\and Osservatorio Astrofisico de Arcetri, Largo Fermi 5, I-50125 Firenze, Italy \\
\email{randich@arcetri.astro.it} 
\and Laboratorio de Astrof\'{\i}sica
 Espacial y F\'{\i}sica Fundamental, INTA, Apartado Postal 50727, E-28080 Madrid, Spain\\ 
\email{barrado@laeff.esa.es}}

\date{Received /Accepted}

\abstract{
We present measurements of the \ion{K}{i} $\lambda 7699$ line from 
spectra of 19 late-type 
members of the $\alpha$Persei cluster obtained 
with the Intermediate Dispersion Spectrograph at the Isaac Newton Telescope. 
These stars span a narrow range of $T_\mathrm{eff}$, from 5091 K to 4771 K, and 
a wide range of $v \sin i$  values (9 $\mathrm{km\,s}^{-1}$--170 $\mathrm{km\,s}^{-1}$). 
We find empirically that larger rotational broadening increases the equivalent width of 
\ion{K}{i} $\lambda 7699$ linearly for $v \sin i$ values from 15 $\mathrm{km\,s}^{-1}$ 
 to 75 $\mathrm{km\,s}^{-1}$. This correlation breaks down for $v \sin i >$  
75 $\mathrm{km\,s}^{-1}$.  After correction for this effect, 
we show that the potassium line equivalent widths do not
correlate with $v \sin i$.}  


\authorrunning{ E.L. Mart{\'\i}n et al.}
\titlerunning{On the potassium-rotation connection}

\maketitle

\section{Introduction}

A long standing problem in stellar evolution has been the connection 
between lithium abundances and rotation in low-mass young stars 
(Soderblom et al. \cite{soder}; Garc{\'\i}a L\'opez et
al. \cite{garcia}). 
Lithium can be destroyed inside stars, and a spread in \ion{Li}{i} 
line strengths could be interpreted
as evidence for a role of rotation in Li depletion. 
Such a role cannot be explained with standard models of lithium 
depletion, simply because they do not include rotation. 
Modifications of the standard model include the effect of the 
 rotational history among stars with the same age and mass (Chaboyer et
al. \cite{chaboy}, Mart{\'\i}n \& Claret \cite{marcla}), magnetic
fields at the base of the convection zone (Ventura et
al. \cite{ventura}) and gravity waves (Montalban \& Schatzman
\cite{monsch}).

An alternative interpretation of the lithium-rotation connection is that it 
is not due to differences in the lithium abundances, but to effects of 
rotation on the formation of the lithium resonance line. Careful analysis 
of the subordinate lithium line at 610.4 nm by Ford, Jeffries \& Smalley (\cite{ford})  
indicates that the spread in lithium abundances among Pleiades stars with 
the same mass may be real. 

Dispersion of \ion{K}{i} $\lambda 7699$ equivalent width measurements
in Pleiades low-mass stars has been reported by Soderblom et
al. (\cite{soder}) and Jeffries (\cite{jeff}). For a given effective
temperature, stars with high chromospheric activity and fast rotation
tend to have stronger \ion{K}{i} lines. 
 The \ion{Li}{i} and \ion{K}{i} lines have similar
formation mechanisms (Stuik et al. \cite{stuik}). Since potassium is
not destroyed in stellar interiors, all cluster members are expected
to have the same abundance. 

The spread in \ion{K}{i} line strength suggests that the dispersion in
\ion{Li}{i} may not be due to abundance differences among the stars,
but to the formation process of the lines. Active stars are known to
have surface spots where the \ion{Li}{i} and \ion{K}{i} lines are
stronger. However, monitoring of the variability of these lines in
active stars in the Pleiades has failed to provide conclusive evidence
for spot-induced modulation (Patterer et al. \cite{pater}, Jeffries
\cite{jeff}). Possible variability on timescales of years has been noted
(Mart{\'\i}n \& Claret \cite{marcla}), which could be due to cycles of
magnetic activity, but this has not been demostrated with long-term
monitoring.  The spread in \ion{K}{i} equivalent widths for stars of
similar $T_\mathrm{eff}$ in the Pleiades cluster remains so far
unaccounted for.

Randich et al. (\cite{randich}) found a similar behaviour for
\ion{Li}{i} in cool stars of  the Alpha Persei open cluster as in their
Pleiades counterparts. We present here \ion{K}{i} equivalent width
measurements in Alpha Persei stars from Randich et
al. (\cite{randich}), with the aim to test whether in these objects
potassium shows a similar spread as lithium. In Sect. 2 we
describe our observations. In Sect. 3 we give an analysis of the
effect of rotational broadening on the measurement of equivalent
widths. We find that the potassium-rotation connection dissapears 
when the equivalent width measurements are corrected for the effect of
rotation on the line profiles. In Sect. 4 we discuss these results.
We conclude that the potassium-rotation connection in late-type members
of young open clusters may largely be due to the effect of rotational broadening 
in the measurement of equivalent widths.

\section{Observations}

Objects of our sample have been selected from Randich et
al. (\cite{randich}). They are listed in Table~\ref{tsample}, together
with their effective temperatures and rotational velocities.  We chose stars 
within a narrow range of $T_\mathrm{eff}$ and spanning a wide range of 
rotational broadening. Optical
spectra were obtained with the Intermediate Dispersion Spectrograph at
the Cassegrain focus of the 2.5m Isaac Newton Telescope in La
Palma. The observations were performed in the night of November 7,
2000. The 235 mm camera, equipped with a EEV CCD, was used in
conjunction with the grating H1800V\@. A resolution of $\sim 1$~{\AA}
was achieved. Exposure times of 1200~s were used for all the spectra,
except in the case of the object AP43, which was observed with an
exposure time of 1600~s. Data reduction, including bias subtraction,
flat-fielding, extraction of one-dimensional spectra, and wavelength
calibration, was carried out using the IRAF\footnote{IRAF is
distributed by the National Optical Astronomy Observatory, which is
operated by the Association of Universities for Research in Astronomy,
Inc., under contract with the National Science Foundation.}
environment.

\begin{table}
\caption{Our sample.}
\label{tsample}
\begin{tabular}{llll}
\hline
\hline
Star       & $T_\mathrm{eff}$ & $v \sin i$ & $W_\lambda$ \ion{K}{i} \\
             &      (K)                    &  ($\mathrm{km\,s}^{-1}$) &
(\AA) \\
\hline
AP33     & 4836  & 9 & 0.38 \\
AP37     & 4977  &  29 & 0.34 \\
AP43     & 4840  &  72 & 0.43 \\
AP56     & 4801   & 110 & 0.41  \\
AP65     & 4798   & 10 & 0.35 \\
AP70     &  4886  &  9 & 0.35  \\
AP72     & 4911   & 9 & 0.31 \\
AP78     & 4775   &  13 & 0.46 \\
AP91     & 4837   & 25 & 0.40 \\
AP93     & 4873   & 75 & 0.50 \\
AP98     & 4912   & 9 & 0.37 \\
AP106   & 4847   & 9 & 0.33 \\
AP110   & 5087   & 9 & 0.33 \\
APX1     & 5091   &  13 & 0.33 \\
APX27A  &  5009  &  50 & 0.45 \\
APX49A  &  4795  &  21 & 0.32  \\
APX59   &  4957   &  38 & 0.34 \\
APX72    & 4893   & 16 & 0.32 \\
APX158   & 5001   & 170 & 0.47\\
\hline
\end{tabular}
\end{table}

\begin{figure}
\includegraphics[width=8cm]{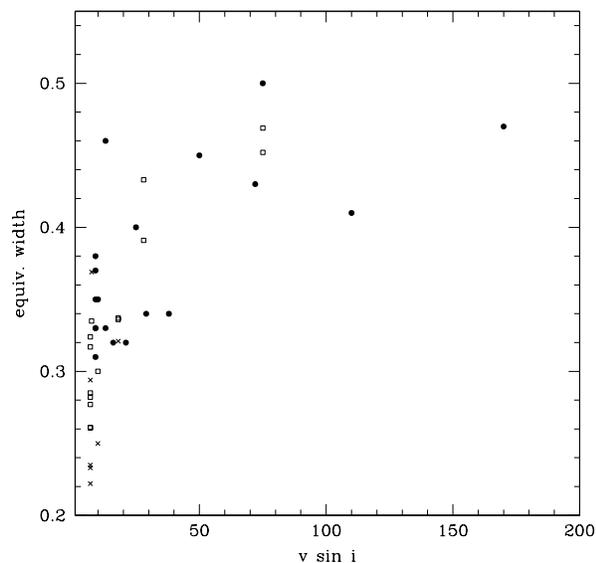}
\caption{Equivalent widths of the \ion{K}{i} $\lambda 7699$ line versus rotation velocity. 
Different symbols denote different samples as follows: our Alpha Persei stars (filled dots), 
Pleiades stars from Jeffries 1999 (open squares) and Pleiades stars from Soderblom et al. 1993 
(crosses). Only stars with $T_\mathrm{eff}$ in the range 5091-4771 K are shown.}
\label{fprreq}
\end{figure}

Equivalent widths of the \ion{K}{i} $\lambda 7699$ line have been measured
by direct integration and are reported in
Table~\ref{tsample}\@. The average 3$\sigma$ error bar is 10\% of the equivalent 
width. For AP33 and AP70 the 3$\sigma$ error bar is 20\% . 
Figure~\ref{fprreq} shows these measurements
vs. $v \sin i$ (dots), together with data obtained in the Pleiades by
Jeffries (\cite{jeff}; open squares) and Soderblom et
al. (\cite{soder}; crosses). This figure indicates a large spread in
equivalent widths, with a trend for stars in the Alpha Persei cluster with higher $v \sin i$ also
to also have a stronger \ion{K}{i} line, consistently with previous results
in the Pleiades. It seems that all stars with $v \sin i$ larger than 30$\mathrm{km\,s}^{-1}$ 
in these two young open clusters have systematically larger  \ion{K}{i} $\lambda 7699$ lines 
than their counterparts with slower rotation rates. This systematic effect is significant 
given the error bars in the equivalent widths ($\le$10\%). However, in the next section 
we will show that the correlation between \ion{K}{i} line and $v \sin i$ is spurious.

It is well known that 
the rotational broadening of a spectrum changes the line profiles and
appearance of the continuum. This can also affect the equivalent
width measurements. Hence, it is not possible to use the same continuum regions 
for all stars in the sample, regardless of their rotational broadening. 
In the next section we develop a method to correct the 
\ion{K}{i} equivalent widths from the effect of rotational broadening.

\section{Analysis}

\subsection {Estimating the effects of rotational broadening}

We broadened the spectra of slow rotators with rotational profiles
of various $v \sin i$'s, measured the equivalent width of the
\ion{K}{i} line at different velocities, and then checked whether such
equivalent widths change with $v \sin i$ in a significant way. In an
attempt to make our test as realistic as possible, for each fast
rotator in our sample we selected low $v \sin i$ ``reference'' stars
in the same sample, according to similarity of the effective
temperature. In Table~\ref{templa} we show the matches made according
to this criterion. The temperature difference between the fast rotator
and reference star is reported as $\Delta
T_\mathrm{eff}$\@.  From the table we see that for each fast rotator
two reference slow rotators have been found with differences in
$T_\mathrm{eff}$ not exceeding 100\,K (average absolute
value $\sim 33~\mathrm{K}$). After the exclusion of AP33 and AP70,
owing to the low signal-to-noise ratio of their spectra, we are left
with five slow rotators which can be used as reference. In
Table~\ref{refer} we report the rotational velocities of the fast
rotators matching these objects in effective temperature.

\begin{table}
\caption{Reference stars for fast rotators.}
\label{templa}
\[ \begin{array}{lll}
\hline
\hline
\mathrm{Fast~Rotator}  & \mathrm{Reference~Star} & \Delta T_\mathrm{eff}
\\

&                                        &   (\mathrm{K})       \\
\hline
\mathrm{AP}37& \left \{\begin{array}[c]{l}
\mathrm{AP}98\\\mathrm{AP}72 \end{array}\right. &\begin{array}[c]{l}
+65\\+66\end{array} \\
\mathrm{AP}43 & \left \{\begin{array}[c]{l} \mathrm{AP}106 \\
\mathrm{AP}33\end{array} \right. &  \begin{array}[c]{l}  -7 \\ +4
\end{array} \\
\mathrm{AP}56 &  \left \{\begin{array}[c]{l} \mathrm{AP}65 \\
\mathrm{AP}33\end{array} \right.  &  \begin{array}[c]{l}   +3 \\-35
\end{array} \\
\mathrm{AP}91& \left \{\begin{array}[c]{l} \mathrm{AP}33 \\
\mathrm{AP}110\end{array}\right.  &  \begin{array}[c]{l}  +1 \\ -10
\end{array}\\
\mathrm{AP}93& \left \{\begin{array}[c]{l} \mathrm{AP}70 \\
\mathrm{AP}106\end{array} \right. & \begin{array}[c]{l} -13 \\ +26
\end{array}\\
\mathrm{APX27A} & \left \{ \begin{array}[c]{l} \mathrm{AP}110 \\
\mathrm{AP}98\end{array}\right. &\begin{array}[c]{l} -78
\\+97\end{array}\\
\mathrm{APX}59 &\left \{\begin{array}[c]{l}\mathrm{AP}98 \\
\mathrm{AP}72\end{array}\right. &\begin{array}[c]{l}+45\\+46
\end{array}  \\
\mathrm{APX}158&\left \{\begin{array}[c]{l}\mathrm{AP}110 \\
\mathrm{AP}98\end{array}\right. & \begin{array}[c]{l} -86 \\ +89
\end{array} \\
\hline
\end{array}  \]
\end{table}

\begin{table}
\caption{Selected reference stars and broadening  velocities.}
\label{refer}
\begin{tabular}{lllll}
\hline
\hline
Star   &   \multicolumn{4}{c}{$v \sin i$}    \\
         &    \multicolumn{4}{c}{(km\,s$^{-1}$)}  \\
\hline
AP65  &  110 & & & \\
AP72  & 29 & 38 & & \\
AP98   & 29   & 38   &  50 & 170 \\
AP106  & 72  &  75 & & \\
AP110 & 25   &  50  &  170   &   \\
\hline
\end{tabular}
\end{table}

\begin{figure}
\includegraphics[width=8cm]{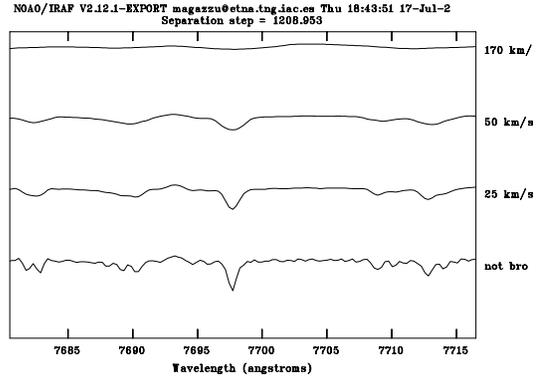}
\caption{The \ion{K}{i} $\lambda 7699$ region of AP110. From bottom to
top: the spectrum ``as it is'', and after convolution with rotational
profiles ($v \sin i = 25, 50, \mathrm{and}~170~\mathrm{km\,s}^{-1}$).}
\label{fbroa}
\end{figure}

A Fortran program written by Ya. V. Pavlenko (2001, private
communication) was used to broaden
the spectra of stars in Table~\ref{refer} by convolution with
rotational profiles of $v \sin i$'s reported in the same table. In
Fig.~\ref{fbroa} we plot a region around the \ion{K}{i} $\lambda 7699$
line. The spectrum of AP110, broadened for several rotational
velocities, is shown.

Equivalent widths of the \ion{K}{i} $\lambda 7699$ line in the spectra
of the reference stars in Table~\ref{refer} have been measured by
direct integration and, when possible, by gaussian fitting, using
tasks within IRAF\@. After broadening, gaussian noise was added. Then,
equivalent widths were measured in the broadened spectra by direct
integration of the line profiles, in the same way as the measurements
performed in the original spectra. The results are shown in
Table~\ref{eqw}\@.

\begin{table*}
\caption{\ion{K}{i} $\lambda 7699$ equivalent widths in the spectra of slow rotators
broadened for various velocities.}
\label{eqw}
\begin{tabular}{ll|lllllllll}
\hline
\hline
Star   & & \multicolumn{8}{c}{\mbox{$W_\lambda$} (\AA)}  \\
         & $v \sin i$  (km\,s$^{-1}$):   & 25  & 29   &  38  &  50 & 72
& 75 & 110 & 170 \\
\hline
AP65   &  &   &    &    &   &    &  & 0.41 &  \\
AP72   &  &  &  0.33 & 0.35 & &  & & & \\
AP98   &  &  & 0.38 & & 0.42  & & & & 0.41  \\
AP106  &  &  & & &   & 0.41 & 0.42  & &   \\
AP110  &  &  0.33  &    & & 0.37  & & & & 0.39  \\
\hline
\end{tabular}
\end{table*}

\begin{figure}
\includegraphics[width=8cm]{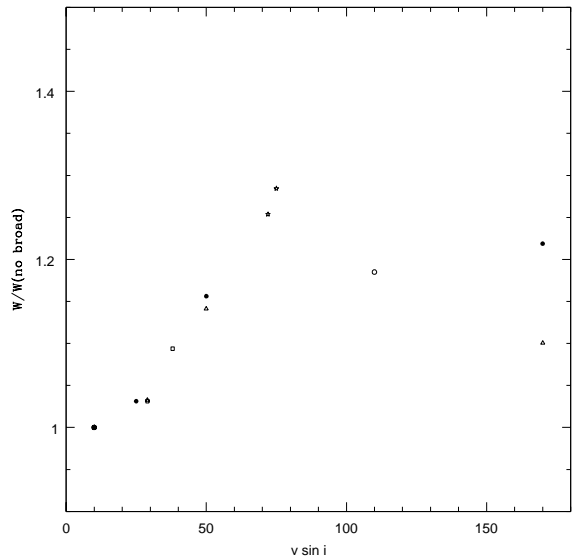}
\caption{$W/W_0$ vs. $v \sin i$.}
\label{fratio}
\end{figure}

For each star in Table~\ref{eqw} we can calculate $W/W_0$, the ratio
between the equivalent width at a given rotational velocity and the
equivalent width in the original spectrum. This quantity is plotted
vs. $v \sin i$ in Fig.~\ref{fratio}\@. In this figure we see that,
while $W/W_0$ increases for $ 15~\mathrm{km\, s}^{-1} \la  v \sin i \la
75~\mathrm{km\, s}^{-1}$, at
100~km\,s$^{-1}$ a small decrease is seen. It is not clear whether at
higher $v \sin i$'s there is a further decrease or if $W/W_0$ can be
considered constant for $v \sin i \ga 100 \mathrm{km\,s}^{-1}$. We can
explain this trend by considering that an increase of the broadening
leads also to an increase of the blending of the \ion{K}{i} line with
adjacent lines, with as a consequence the enlargement of the
measured equivalent width. However, for higher velocities, the lines
become so shallow that the integration extremes are difficult to
define, which will tend to underestimate the integration
range. Therefore, a fraction of flux does not contribute to the
equivalent width, somewhat compensating for the blending. The velocity
corresponding to the breakdown of the linear increase depends on the
resolution of the data.

In any case, Figure~\ref{fratio} indicates a significant increase of
the equivalent width due to rotational broadening.  Data in
Fig.~\ref{fratio} fit a very tight linear relationship (correlation
factor = 0.996) for $15~\mathrm{km\,s}^{-1} < v \sin i <
75~\mathrm{km\,s}^{-1}$. For faster velocities, we suppose a constant
behaviour and set $W/W_0$ equal to the average of the three highest
velocity points in our sample. For $v \sin i < 15~\mathrm{km\,s}^{-1}$
we cannot calibrate the effect of rotation because of the modest
resolution of our spectra. In summary:

\begin{equation}
\label{eq1}
W/W_0 = \left \{ \begin{array}{ll}
0.89 + 5.14 \times 10^{-3} \cdot v \sin i \hspace{2 ex} & \mathrm{if}\,
15~\mathrm{km\,s}^{-1} < v \sin i < 75~\mathrm{km\,s}^{-1} \\
1.17   & \mathrm{if}\, v \sin i > 75~\mathrm{km\,s}^{-1} 
\end{array} \right.
\end{equation}

\subsection{Correcting our equivalent widths}

\begin{table}
\caption{Corrected equivalent widths}
\label{tfineq}
\begin{tabular}{ll}
\hline
\hline
Star       & $W_\lambda$ \\
             &   (\AA)      \\
\hline
AP37    & 0.34 \\
AP43    & 0.43 \\
AP56    & 0.41  \\
AP91    & 0.40 \\
AP93    & 0.50 \\
APX27A  & 0.45 \\
APX49A  & 0.32  \\
APX59   & 0.34 \\
APX72   & 0.32 \\
APX158  & 0.47\\
\hline
\end{tabular}
\end{table}

We use these results to correct our measurements. According to
Eq.\ref{eq1}, we determine $W/W_0$ for each of the stars observed in
this work and divide the equivalent widths in Table~\ref{tsample} by
this quantity. This allows us to correct our equivalent widths for the
effects of rotational broadening. Final equivalent widths are
listed in Table~\ref{tfineq} and shown vs. $v \sin i$ in
Fig.~\ref{ffineq}\@.

\section{Discussion and perspectives}

\begin{figure}
\includegraphics[width=8cm]{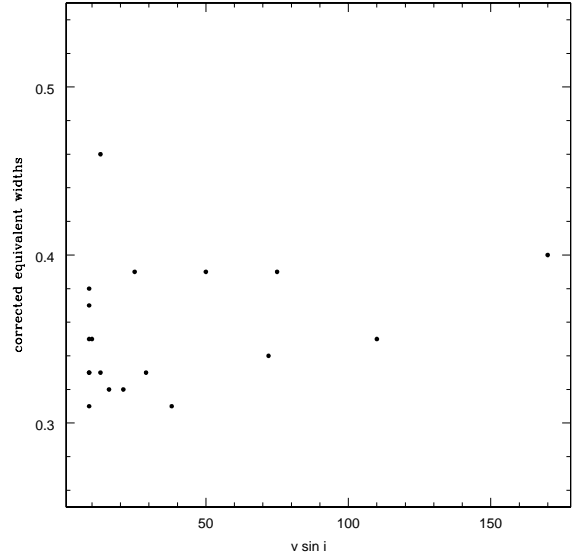}
\caption{Corrected equivalent widths vs. $v \sin i$.}
\label{ffineq}
\end{figure}

A look at Fig.~\ref{ffineq} suggests no significant correlation
between \ion{K}{i} equivalent widths and $v \sin i$. This is
corroborated by the poor correlation factor, equal to 0.26\@. Note
that the corresponding correlation factor for data in
Table~\ref{tsample} (equivalent widths before correction) is
0.64\@. We see also that the average corrected equivalent width for
stars with $v \sin i \le 15~\mathrm{km\, s}^{-1}$ is $0.36 \pm
0.4$~\AA, practically the same as for faster rotators ($0.35 \pm
0.4$~\AA).

Thus, in our Alpha Persei sample, the faster rotators do not show
stronger \ion{K}{i} $\lambda 7699$ lines than the slow rotators. The
increase of equivalent width with $v \sin i$ can be explained in terms of a systematic
effect due to rotational broadening. The effect found in this work
could also be present in other data sets of similar resolution. This
could also explain the trend seen in the Pleiades (Soderblom et
al. \cite{soder}, Jeffries \cite{jeff}). 

There is still a significant spread in the \ion{K}{i} equivalent widths shown 
in Fig.~\ref{ffineq}. It seems that the effect of $v \sin i$ on the measurement 
of equivalent widths does not explain completely the scatter in the line 
strengths. Other effects may be present that change the formation of the lines, 
but they do not depend on rotation. Possibly the spread is caused by small 
differences in $T_\mathrm{eff}$ among the stars in our sample because the 
line is very sensitive to temperature. For example, the star with 
the largest \ion{K}{i} equivalent width (AP78) in Fig.~\ref{ffineq} 
is also the coolest star included in our study ($T_\mathrm{eff}$=4775~K).  
Another factor that could play a role in the spread of corrected  \ion{K}{i} equivalent widths 
is stellar activity because of the effect of stellar spots and faculae on the line 
formation (Barrado y Navascu\'es et al. \cite{barrado}). 

Since the \ion{K}{i} $\lambda 7699$ line is considered to be a good proxy of 
the \ion{Li}{i} $\lambda 6708$ line, we plan to perform a similar test of 
the dependence of the \ion{Li}{i} equivalent width on $v \sin i$ in a future work. 
The assumption that the equivalent widths of individual lines is invariant for 
different values of $v \sin i$ may need to be revised.

\end{document}